\documentstyle[aps,multicol,prl]{revtex}

\begin{document}
\title{Slow fluctuations in enhanced Raman scattering and surface roughness
relaxation }
\author{D. B. Lukatsky$^{a}$\footnote{Present address: FOM Institute for Atomic and 
Molecular Physics, Kruislaan 407, 1098 SJ Amsterdam, The Netherlands}, G. Haran$^{b}$, and S. A. Safran$^{a}$}
\address{$^{a}$Department of Materials and Interfaces, and $^{b}$Department
of Chemical Physics,\\
Weizmann Institute of Science, 76100 Rehovot, Israel}
\maketitle

\begin{abstract}
We propose an explanation for the recently measured {\it slow fluctuations}
and ``blinking'' in the surface enhanced Raman scattering (SERS) spectrum of
single molecules adsorbed on a silver colloidal particle. We suggest that
these fluctuations may be related to the dynamic relaxation of the surface
roughness on the nanometer scale and show that there are two classes of
roughness with qualitatively different dynamics.\ \ The predictions agree
with measurements of surface roughness relaxation. Using a theoretical model
for the kinetics of surface roughness relaxation in the presence of charges
and optical electrical fields, we predict that the high-frequency
electromagnetic field increases both the effective surface tension and the
surface diffusion constant and thus accelerates the surface smoothing
kinetics and time scale of the Raman fluctuations in manner that is linear
with the laser power intensity, while the addition of salt retards the
surface relaxation kinetics and increases the time scale of the
fluctuations. These predictions are in qualitative agreement with the Raman
experiments.
\end{abstract}


\begin{multicols}{2}

In\ a recent experiment by Weiss and Haran \cite{Haran}, large spectral
fluctuations in the relative intensities of different Raman lines that
varied on a time scale of a few tens of seconds, were measured in the
surface-enhanced Raman scattering (SERS) of single rhodamine 6G molecules
adsorbed on silver nanocrystals. The rate of spectral fluctuations was
demonstrated to {\it increase} with laser intensity and {\it decrease} with
addition of salt in the solution of the silver colloidal particles. In
addition, a decay of the overall intensity of the scattering was observed on
a scale of the order of hundreds of seconds; this decay was also correlated
with the laser intensity \cite{Haran}. Finally, fluctuations of the overall
intensity on a time scale comparable with that of the fluctuations of the
individual spectral lines were measured in Refs. \cite{Haran}. These
fluctuations, sometimes termed ``blinking'', are frequently observed in
single-molecule SERS studies \cite{Nie97,Nie02,Michaels99,Bjerneld00}. The
long time scale observed in the modulations of the spectrum of molecules
adsorbed on silver colloidal particles, was suggested \cite{Haran} to arise
from (slow) motion of the adsorbed molecule, which leads to variation of a
charge-transfer interaction between the molecule and the surface.

In this paper, we propose an alternative interpretation of both the slow
SERS spectrum fluctuations, and of the decay and the fluctuations of the
overall intensity (``blinking'')\ of the SERS spectrum measured in Ref. \cite%
{Haran} and show that the time scales for these phenomena are consistent
with the relaxation of the surface roughness on the nanometer scale \cite{fn}. It is
well known that a significant part of the enhancement of the SERS signal is
due to the surface roughness \cite{Moskovits85,Otto83}. Using theoretical
models for the kinetics of surface roughness relaxation in the presence of
charges and optical electrical fields, we predict the dependence of the time
scale of the fluctuations on the laser power intensity and on the amount of
added salt in agreement with the observations.

Conventionally, the SERS enhancement mechanism is separated into 
the electromagnetic (EM) field enhancement and the chemical enhancement
\cite{Nie02}. The relaxation of surface roughness can affect both the EM 
and chemical mechanism of the SERS enhancement, depending on the 
characteristic length-scale and the amplitude of the relaxation modes.
The surface relaxation dynamics itself, however, is governed 
by the EM mechanism in our model, as we elaborate below. These
constitute the starting point of our analysis.

Our dynamical model applies both to equilibrium fluctuations and to the
decay to equilibrium (a smooth surface) of non-equilibrium, surface
roughness that can have large amplitudes, depending on the initial
preparation of the nanoparticles. The amplitude of equilibrium fluctuations
is typically small for perfectly smooth surfaces \cite{Sambook}.\ \ However,
rough surfaces (where the roughness due to sample preparation decays slowly)
show larger thermal fluctuations \cite{roughsurface}. \ Thus, the roughness
fluctuations may have a strong effect on the SERS enhancement mechanism and
may lead to the observed fluctuations in the relative intensities of the
different Raman lines.

We first estimate the time scale for relaxation of the surface roughness. We
use a model for surface relaxation developed by Mullins \cite{Mullins57}
based on an isotropic expression for the surface energy. This theory is
applicable either above the roughening transition temperature or for vicinal
surfaces (that are not perfectly smooth on the atomic scale), even below the
roughening temperature. \ The dynamics of ideal, high symmetry surfaces
involves the creation of steps and facets \cite{kandel} and is outside the
scope of our work. Assuming that surface diffusion (with diffusion constant $%
D_{s}$ for single atom motion on the surface) is the only relevant process,
leads one to the kinetic equation:
\begin{equation}
\frac{\partial h}{\partial t}=-\Gamma \,\nabla ^{4}h\,,  \label{kinetics1}
\end{equation}%
where $h(\vec{\rho},t)$ is the local height, $\gamma _{0}$ is the surface
tension, $\vec{\rho}=(x,y)$ is the in-plane position vector and $\vec{\nabla}%
=(\partial _{x},\partial _{y})$, and $\Gamma =D_{s}a^{4}\gamma
_{0}\,/\,k_{B}T$ with $a$ being an atomic length scale (e.g. the
nearest-neighbor distance, for silver $a\simeq 2.8$ \AA ). To estimate the
characteristic time of the SERS spectral fluctuations, we analyze the
height-height dynamic correlation function, ${\cal C}_{hh}(\vec{\rho}%
,t)\equiv \left\langle h(\vec{\rho},t)h(0,0)\right\rangle \,$.

${\cal C}_{hh}(\vec{\rho},t)$ can be straightforwardly obtained from Eq. (%
\ref{kinetics1}) (see e.g., Refs. \cite{Sambook,Lubensky}) assuming that the
surface atoms are in contact with a thermal bath: ${\cal C}_{hh}(\vec{\rho}%
,t)=\frac{k_{B}T}{2\pi \gamma _{0}}\int_{q_{0}}^{\infty }\frac{e^{-\Gamma
q^{4}t}}{q}J_{0}(q\rho )\,$d$q$, where $J_{0}(x)$ is the Bessel function,
and $q_{0}=2\pi /L$, with $L$ being the system size. To estimate the
characteristic relaxation time, $\tau \,$, numerically, one can adopt the
criterion: ${\cal C}_{hh}(\vec{\rho}_{0},\tau )/{\cal C}_{hh}(\vec{\rho}%
_{0},0)=1/\exp(1)$, were $\rho _{0}$ is of the order of a size of the
adsorbed molecule.

An accurate estimate of the relaxation time, $\tau $, requires a knowledge
of the surface diffusion constant, $D_{s}$, of silver estimated in the range
$D_{s}\simeq 1.8$ $\times \,10^{-15}$ cm$^{2}/\sec $ \ \cite{Pai97} to $%
D_{s}\sim 10^{-14}$ cm$^{2}/\sec $ \ \cite{Hirai98}. Using $L\simeq 50$ nm
of the order of a size of silver colloids in the experiment \cite{Haran},
surface tension of Ag \cite{Adamson},\ $\gamma _{0}\simeq 1500$ erg/cm$^{2}$%
, $\rho _{0}\simeq 6$ \AA , we find that the correlation time of the equilibrium
fluctuations is $\tau \simeq 18$ sec. This shows that the time scale for the%
{\it \ equilibrium} relaxation of surface roughness on nanometer scale at
room temperature is comparable to the correlation times of the slow
fluctuations of the SERS\ spectrum measured in \cite{Haran}. An analytic
estimate for the correlation time can be obtained from the asymptotic
long-time form: ${\cal C}_{hh}(\vec{\rho}_{0},t)\simeq {\cal C}%
_{hh}(0,t)\sim \frac{k_{B}T}{\gamma _{0}}\,\frac{\exp (-t/\tau )}{t/\tau }\,$%
, where $\tau =1/\Gamma q_{0}^{4}$. We emphasize that the relaxation time, $%
\tau \,$, is sensitive to the numerical values used for the lattice constant
as well as to the wavelength associated with the size of a colloid.

In addition to the small amplitude, equilibrium fluctuations of the surface,
much larger amplitude surface roughness may arise due to the sample
preparation. These features tend to decay in time, leading to a smooth
surface at the atomic level and this relaxation to equilibrium of a
relatively smooth surface (and hence less enhancement of the Raman
scattering), leads to the overall decay of the SERS spectrum intensity. The
loss of the Raman signal is a well known phenomenon that also occurs in
electrochemical \cite{SERSlost1,Dick02} and ultrahigh vacuum \cite%
{SERSlost2,Douketis00} SERS systems. This effect was attributed to a
diffusive loss of surface adatoms in a number of investigations (see e.g.,
introduction in Refs. \cite{SERSlost1,Dick02}); however, none of these, to
the best of our knowledge, explain the slow time scale in terms of
cooperative surface tension effects nor do they discuss the dependence of
the decay on the laser optical field and on added salt.

Another type of experiment to which our kinetic model is relevant, directly
measures the dynamics of artificially created, nanoscale surface roughness
features \cite{Hirai98}. Since the surface profiles in the experimental
systems are, in general, far from being simple sinusoids, we predict the
relaxation time for two important classes of surface features.

To estimate the relaxation time for the two classes of surface features we
consider two specific surface roughness profiles at an initial time $t=0$:
(i) a non-mass-conserving profile, modelled as a Gaussian protrusion where $%
h>0$, $h(\vec{\rho},t=0)=h_{0}\,e^{-\alpha \rho ^{2}}$ and (ii) a
mass-conserving profile, modelled as a region where a protrusion with $h>0$
is adjacent to an indentation where $h<0$. The average value of $h$ over the
entire surface is zero and for convenience we consider: $h(\vec{\rho}%
,t=0)=h_{0}\,e^{-\alpha \rho ^{2}}(1-\alpha \rho ^{2})$, where $\alpha
^{-1/2}$ is the characteristic lateral scale of the feature, and $h_{0}$ is
the amplitude. The experiments of Ref. \cite{Hirai98} correspond to the
mass-conserving case (ii) (possibly because of the manner in which the
surface was scratched), while we expect that general surface roughness of
colloidal particles, applicable to the Raman experiments, to be more similar
to the non-mass-conserving case (i). It is straightforward to compute the
time evolution of these two types of profiles, by solving Eq. (\ref%
{kinetics1}) with the corresponding initial conditions of cases (i) and
(ii). The time dependence of the decay of the maximum height (located at the
origin, $\rho =0$) of these Gaussian-like peaks can be obtained
analytically. For case (i), we find $h(0,t)=h_{0}\,(\pi \tau
_{g}/t)^{1/2}\,\beta (\tau _{g}/t)\,.$ For case (ii) we find $h(0,t)=h_{0}\,%
\left[ 2\,\tau _{g}/t-\,2\sqrt{\pi }(\tau _{g}/t)^{3/2}\,\beta (\tau
_{g}/t)\,\right] ,$ where $\beta (x)\!=\!\exp (x)$erfc$(\sqrt{x})$ and $\tau
_{g}\!=\!k_{B}T/(64\gamma D_{s}\alpha ^{2}a^{4})$ is the characteristic
decay time, and erfc($x$) is the complementary error function.

The most important observation is that these two types of profiles have
qualitatively different smoothing kinetics -- the mass conserving profile
(ii) decays much {\it faster} since in this case the transport of matter
need only occur near the boundary between the protrusion and the
indentation; that is, atoms are locally transferred from the region where $%
h>0$ to the region where $h<0$. For case (i) of the non-mass-conserving
profile, the matter must be transported to a much larger scale. This is
reflected in the expressions for the asymptotic, long time ($\tau _{g}\ll t$%
) evolution of the height. \ For case (i), the decay of the maximum of the
profile (located at the origin), $h(0,t)\simeq h_{0}\,\pi ^{1/2}\,(\tau
_{g}/t)^{1/2}$. For case (ii) $h(0,t)\simeq 2\,\,h_{0}\,\tau _{g}/t\,$ tends
to zero much faster.

We now use this model for surface smoothing in case (ii) to estimate the
decay time for the (approximately) mass-conserving surface features studied
in Ref. \cite{Hirai98}. Using their estimate for the diffusion constant at
zero external potential, $D_{s}\simeq 10^{-14}$ cm$^{2}/\sec $ and a value
for $\alpha ^{-1/2}\simeq 8$ nm that corresponds to the extent of the
scratch, we obtain that a mass conserving, Gaussian profile with initial
amplitude $h_{0}\simeq 2$ nm decays to an atomic-scale estimated as $h(\tau
)\simeq 0.3$ nm, in a time $\tau \simeq 200$ sec. This theoretical estimate
is consistent with the experimentally measured times in the scratch
experiments of Hirai \cite{Hirai98}. An estimate of the decay time for case
(i) (i.e., the non-mass-conserving profile), yields for the same parameters,
a decay time that is about 15 times slower, $\tau \simeq 3100$ sec. In the
case of colloidal particles, we expect that typical, non-equilibrium
features (preparation dependent roughness) may be non-mass-conserving
protrusions or indentations and that the kinetics of case (i) would apply.
We estimate, for instance, that a more localized, Gaussian protrusion of
amplitude $1$ nm and extent $\alpha ^{-1/2}\,\simeq 6\,$nm would decay to an
atomic size of $0.3$ nm in a time $200$ sec, consistent with the times
measured in the Raman experiments \cite{Haran}.

The Raman experiments show a systematic dependence of the modulation and
relaxation time scales on the salt and the electric field as mentioned at
the beginning. \ Since the characteristic time scales vary inversely with
the product of the surface tension and diffusion constant (see Eq. (1)), we
consider:\ (i) the effect of salt on the surface tension (ii) the effect of
the laser field on the time scale via its effect on the surface tension
(iii) the effect of the laser field on the surface diffusion constant. To
treat the effects of surface charges \cite{fnSurfaceCharges} and the laser
electric field on the silver surface tension and hence on the surface
relaxation, we model the charged colloidal interface as an elastic, almost
planar surface, with a fixed and uniform surface density of charge $\sigma $%
, and height $h(\vec{\rho}\,)$. \ The surface tension, given by calculating
the free energy cost (including the electrostatic effects due to the charges
and the salt) of deviations of the surface from the planar geometry \cite%
{Goldstein90} is: $\gamma _{e}=\frac{2\pi \sigma ^{2}}{\epsilon \kappa }\,=%
\frac{\epsilon E_{s}^{2}}{8\pi \,\kappa }\,$, where $\epsilon \simeq 80$ is
the dielectric constant of water, $\kappa ^{-1}$ is the Debye screening
length, $\kappa ^{2}=8\pi \ell _{B}{\cal Z}^{2}n$, here $\ell _{B}=\frac{%
e^{2}}{\epsilon k_{B}T}\simeq 7$\AA\ is the valence of salt ions, and $n$ is
the salt concentration, $E_{s}=4\pi \sigma /\epsilon $ is the electric field
at the colloid surface. This result can also be obtained from a scaling
argument: the tension is the product of the energy density and the volume
divided by the cross-sectional area; this is proportional to energy density $%
\epsilon E^{2}$ multiplied by a characterisitic length, which here is the
Debye length, $\kappa ^{-1}$.

This result shows that with the addition of salt, the effective surface
tension {\it decreases, }$\gamma _{e}\sim n^{-1/2}$. This leads to an {\it %
increase} of the relaxation time, $\tau $, as observed. Taking the
experimental values of the parameters used above, we see that in order to
obtain an effect of order unity (i.e., an effective electrostatic surface
tension, $\gamma _{e}$\thinspace , equal to the bare surface tension of
silver, $\gamma _{e}=\gamma _{0}\simeq 1500$ erg/cm$^{2}$) the surface of a
silver colloidal particle must have a surface charge density, $\sigma \simeq
10\,e/$nm$^{2}$ in $100$ mM salt solution. This is a rather high surface
charge density, but is still within a realistic range for the experiments.

To quantify the contribution of the laser field to the surface energy and
diffusion constant of a silver nanocrystal requires an accurate value for
the enhanced surface electric field; this requires an accurate model of the
microscopic mechanism of the enhancement, which is not yet completely
understood \cite{Moskovits85}. However, we can predict the functional
dependence of the both the surface tension as well as the surface diffusion
constant on the field at the surface. The surface energy density, $u$, of a
semi infinite metal sample with a planar surface boundary in the presence of
a high-frequency, optical electromagnetic field, is \cite{Landau}:
\begin{equation}
u(z\,)=\frac{1}{8\pi }\left[ \frac{\partial (\omega \varepsilon (\omega ))}{%
\partial \omega }\,|\vec{E}(z)|^{2}+\frac{\partial (\omega \mu (\omega ))}{%
\partial \omega }\,|\vec{H}(z)|^{2}\right] \,,  \label{energy1}
\end{equation}%
where $\vec{E}(z)=$ $\vec{E}_{0}\,e^{i\,k\,z}$ is the spatially-dependent
part of the electric field vector with amplitude, $\vec{E}_{0}$, and the
plane, $z=0$, corresponds to the metal interface plane; the same definitions
apply to the magnetic field, $\vec{H}$. For the frequency-dependent
dielectric function, $\varepsilon (\omega )$, we use the simplest dispersion
model of the free, classical electron gas \cite{Landau}: $\varepsilon
(\omega )=1-(\frac{\omega _{p}}{\omega })^{2}$, where $\omega _{p}\simeq 9.2$
eV$\simeq 1.4\times 10^{16}$ sec$^{-1}$ is the plasma frequency of silver,
and $\omega \simeq 3.5\times 10^{15}$ sec$^{-1}$ for the $530$ nm
wavelength laser beam. For the optical frequencies relevant to the
experiments \cite{Haran}, $\mu (\omega )=1$. Using the dispersion relation %
\cite{Landau}, $k^{2}=\frac{\varepsilon (\omega )}{c^{2}}\,\omega ^{2}$, we
obtain: $k\equiv i\chi =i\frac{\sqrt{\omega _{p}^{2}-\omega ^{2}}}{c}$, and
therefore, both the electric and magnetic fields decay exponentially within
the silver colloidal particle with a typical decay length \cite{bookcluster}%
, $\chi ^{-1}$, of $22$ nm. In this frequency region therefore, the metal is
reflecting with a skin depth, $\chi ^{-1}$. Using the relationship between $%
\vec{H}$ and $\vec{E}$ \cite{Landau}, we obtain the time-averaged energy per
unit surface area ({\it i.e., }effective surface tension), $\gamma
_{em}=\int_{0}^{\infty }u(z)\,dz$, in the presence of an electromagnetic
field: $\gamma _{em}=\frac{E_{0}^{2}}{8\pi \,\chi }\,$. Again, this can be
predicted by a scaling argument: the tension is the product of the energy
density (proportional to the laser intensity) and the characteristic length,
here the skin depth, in qualitative agreement with the experiment.

However, an estimate of the magnitude of this effect using the {\it bare }%
values $E_{0}\simeq 0.6$(erg$^{1/2}$/cm$^{3/2}$)$\,\simeq 180$ V/cm,
(corresponding to a laser power density of $100$ W/cm$^{2}$) and $\chi
^{-1}\simeq 22\,$nm is $\gamma _{em}\simeq 3\times 10^{-8}\,\frac{\text{erg}%
}{\text{cm}^{2}}$. This quantity, $\gamma _{em}$, is thus about $10^{11}$
times smaller than the bare coefficient of surface tension of silver, $%
\gamma _{0}$ ! It is known that the electric field at the surface in the
case of SERS is greatly enhanced by the surface plasmon resonances in the
system \cite{Moskovits85,Nitzan82}. Recent experiments \cite{Nie97,Nie02}
report that the SERS enhancement factor, (theoretically predicted to be
proportional to $({\cal E}/E_{0})^{4})$ where ${\cal E}$ is the intensity of
scattered, enhanced optical field (see also Refs. \cite{Xu00,Corni02})),
reaches a value of $10^{14}-10^{15}$. This enhancement is still not large
enough to bridge the gap between our prediction for the optical
field-induced effective surface tension, $\gamma _{em}$, and the bare
surface tension of silver, $\gamma _{0}$. We note, however, that the
intensity of the localized field at the surface of a metal colloid may
exceed the intensity of the radiated, enhanced optical field \cite%
{Moskovits85}. Atomic scale, globally distributed roughness (adatoms,
terraces, kinks, small islands) may produce an extra enhancement of the
electric field at the surface of a colloidal particle, as well, although a
quantitaive understanding of this mechanism is still lacking (see e.g., \cite%
{Otto01}). In summary, the scaling of the tension and hence the time scale
for surface relaxation with the electric field is qualitatively correct;
however, the magnitude depends on the actual value of the surface field. \

Although\ the effect of the field on the surface tension may turn out to be
too small to matter, the change in the local diffusion constant of silver
adatoms with field may be more significant; the diffusion constant enters
the kinetic equations discussed above. Indeed, in the case of a {\it static}
electric field applied to a roughened metal interface, Hirai et al. \cite%
{Hirai98} show from their experiment that the surface diffusion coefficient,
$D_{s}$, depends {\it exponentially} on the applied potential difference.
This is in accord with theoretical calculations \cite{bias00} of surface
diffusion in the presence of external fields, in the limit where the
electrostatic energy is much larger than the thermal energy. This is the
case in the static experiments where the potential drop occurs on a length
scale of $\sim 1$ nm of order of the Debye length; this gives rise to a very
large electric field. However, this limit is not applicable to the Raman
experiments. The electrolyte cannot respond to the high frequency optical
field and provides no screening of the laser field. \ The potential drops
over a length scale given by the optical wavelength, and not the Debye
length. The resulting field is thus orders of magnitude smaller than in the
static case. Indeed, the Raman experiments show only a linear dependence of
the rate of the SERS spectral fluctuations on the laser intensity and no
exponential behavior.

Calculations \cite{bias00} of surface diffusion in the presence of external
fields in the limit where the field energy is smaller than the thermal
energy, show that the surface diffusion coefficient depends {\it linearly}
on the laser intensity \cite{Noise}: $D_{s}\simeq D_{s}(E_{0}=0)\left(
1+W^{2}\right)$ where $W\equiv eE_{0}a/(4k_{B}T)\ll 1$. This trend agrees
with the experimental observation that the time scale varies with the laser
intensity. However, as in the case of the surface tension, this requires a
significant enhancement of the local, surface field to yield a measureable
effect. We note that even if the field contribution to the tension is small,
the effect on the diffusion constant may be significantly larger since the
field dependence of the tension depends on the ratio of the intensity to the
bare silver surface tension while the correction to the diffusion constant
varies with the ratio of the intensity to the thermal energy, which is about
two orders of magnitude smaller than the energy associated with the surface
tension of silver.

The best way to verify our predictions would be to perform SERS measurements 
on surfaces with \textit{in situ} control of surface roughness by \textit{e.g.}, 
the methods described in Ref. \cite{Hirai98}. Monitoring the SERS spectrum as a 
function of the surface roughness, the salt concentration, and the laser intensity
should provide an ultimate test of our idea.

We thank J. Imry, D. Kandel, and A. Nitzan for useful discussions. SAS acknowledges the
support of the US Israel Binational Science Foundation and the Schmidt
Minerva Center. GH is the incumbent of the Benjamin H. Swig nad Jack D.
Weiler career development chair.


\end{multicols}


\begin{references}
\bibitem{fn} We note that in the single-molecule SERS experiments in 
addition to intensity fluctuations, frequency fluctuations are usually 
observed (see e.g., Ref. \cite{Nie02}). These fluctuations were much less 
pronounced compared with the intensity fluctuations in Ref. \cite{Haran}.
These large, intensity fluctuations are the focus of our work.
\bibitem{roughsurface} One can see this by expanding the full expression for
the surface energy, $\gamma \int \text{d}\vec{\rho }\,\,\left[ 1+\left(
\nabla h\right) ^{2}\right] ^{1/2}$ , to second order about the average
roughness and calculating the amplitude of thermal fluctuations.
\bibitem{fnSurfaceCharges} The colloid surface is charged due to the
presence of citrate ions in the solution.
\bibitem{Noise} This result was derived under the assumption that the laser
frequency is smaller than the frequency of the thermal noise. The thermal
noise arising from phonons has a typical frequency of $10^{13}$ sec$^{-1}$
which is about two orders of magnitude smaller than the laser frequency.
However, very high frequency thermal noise may arise from the coupling of
the silver ions to the surface plasmons. \
\end{references}
\end{document}